\documentclass{cernrep}
\begin{document}
\title{Cosmic ray induced micro black hole showers}
\author{M.V. Garzelli, M. O'Loughlin and S. Nafooshe
}
\institute{University of Nova Gorica, Laboratory for Astroparticle Physics, SI-5000 Nova Gorica, Slovenia}
\maketitle

\begin{abstract}
Extended air showers originate from interactions between ultra-high-energy 
cosmic rays and nuclei in the Earth's atmosphere. At present there are some discrepancies between experimental observed properties of these air showers and theoretical predictions obtained by using standard hadronic interaction models for cosmic ray primaries with laboratory energies above 10$^{5} - $10$^{6}$  TeV. In this contribution, we will present a preliminary discussion of the possibility (in the framework of TeV gravity models) that shower development may begin with the production of a microscopic black hole (MBH) at the moment of the primary collision, which then evaporates and decays, by emitting gravitons and Standard Model quanta. From our preliminary investigations it appears that lepton distributions are more likely to reveal the presence of a MBH than photon distributions.
\end{abstract}


When making a comparison between coupling constants related to different types of interaction, one finds $G_{grav}/G_{Fermi} \sim 10^{-34}$ so that $G_{grav} << G_{Fermi}$. In terms of the corresponding enegy scales this means that the 
Planck scale $\Lambda_{Planck } \sim 10^{16}$ TeV, at which gravity becomes strongly coupled and its effects cannot be neglected any more, is far larger than the electroweak (EW) symmetry breaking scale $\Lambda_{EW} \sim 0.25$ TeV. The absence of a clear explanation for a difference of so many orders of magnitude, is a manifestation of the so-called hierarchy problem. A possible solution, dating back to 1998~\cite{arkani-hamed,antoniadis,randall}, proposes  the existence of $n$ extra-dimensions where gravitational interactions can extend (the ``bulk''), thus diluting their effect in our 4-dimensional world (the ``brane'') where the strong and electroweak interactions would be confined. 
Thus, beside the $\Lambda_{Planck }$ scale related to the strength of gravitational interactions in our 4-dimensional world, i.e. $\Lambda_4 = \Lambda_{Planck}$, a fundamental gravity scale in $D$-dimensions ($D = 4 + n$) is introduced which may be as low as $\Lambda_D \sim \Lambda_{EW}$ and possibly also lead to the unification of the fundamental forces. If this would be the case then the effects of gravity would begin to manifest themselves already at this (low) unification energy scale and thus would be within reach of terrestrial accelerator experiments. In particular, one of the most intriguing phenomenological consequences of this scenario would be the possible formation of microscopic black-holes (MBH)'s in collisions between two hadrons with a Center-of-Mass (CM) energy as low as $E_{CM}\sim$ 5 $-$ 20 TeV.  

This contribution deals with hadronic collisions in a wide interval of energies,
ranging from LHC energies ($E_{CM} \sim 10$ TeV) up to the highest cosmic
ray energies ($E_{lab} \sim 10^{6} - 10^{8}$ TeV, equivalent to 
$E_{CM} \sim 40 - 140$ TeV). In this energy range gravitational effects are
usually neglected. These energies are higher than those tipically covered by 
nuclear physics studies\footnote{as those discussed in many contributions to this Conference ($13^{th}$ International Conference on Nuclear Reaction Mechanisms, Varenna, Italy, June 11 - 15 2012)}, however it is possible to establish some parallelism between concepts familiar to nuclear physicists, which determine the evolution of an excited nuclear system formed in the collision between two nuclei at energies of the order of hundreds MeV/nucleon, and concepts that characterize the evolution of a MBH formed by the collision of two hadrons or nuclei at much higher energies as predicted by the Hoop Conjecture~\cite{thorne:1972}: {\it if in the collision of two hadronic objects a large amount of energy/mass is concentrated in a spatial region that can be surrounded by a hoop with a radius $R < R_{Schwarzschild}$ corresponding to a Schwarzschild black hole of that energy, then a MBH is formed.}

Although the hoop conjecture provides some basic necessary conditions for the formation of a MBH in collisions, the actual formation process and the initial phase of its evolution is a very complex non-linear phenomenon, subject to many uncertainties, evolving from an asymmetric configuration out of thermal equilibrium, to a highly symmetric static configuration with a well-defined Hawking temperature. Numerical Monte Carlo simulations may be the best way to model this type of process. The same happens when describing the formation of an excited nuclear system from the collision of two nuclei. In both cases the key parameters determining the evolution of the system are $E_{CM}$ and the total angular momentum $J$, related to the impact parameter $b$ between the initial colliding objects.

At the end of the dynamical phase, it is commonly believed that the MBH undergoes a Hawking evaporation phase during which its temperature evolves following the law $T \propto k/M_{MBH}$, and this can be described by statistical/thermodynamical models together with corrections to the trajectory of the emitted particles as a consequence of the curved geometry through which they subsequently propagate via the so-called grey-body factors. Analogously, the evaporation of nucleons by excited nuclei at the end of the pre-equilibrium phase is commonly described by statistical methods.  

In particular the dynamical + statistical MBH evolution is commonly divided in four sequential phases (for a more detailed discussion see e.g. Ref~\cite{park} and references therein):
\begin{itemize}
\item {\it Balding phase}: the MBH just formed, initially characterized by a deformed shape,``loses its hairs'' (i.e. the higher angular momentum powers), by emitting charge, energy and angular momentum in the form of gravitational radiation and gauge fields, becoming more symmetric (elliptical) in the process.

\item {\it Spin-down phase}: the now stationary rotating MBH continues to gradually lose its energy/mass (60 $-$ 80\%) and angular momentum, until it reaches a non-rotating static (spherical) configuration.

\item {\it Schwarzschild/Evaporation phase}: the MBH loses its mass by emitting all possible particle degrees of freedom (Standard Model particles: quarks, leptons, photons, $W$, $Z$, gluons + gravitons, and, if they exist, other heavy particles beyond the SM). At this stage, the emission is assumed to be democratic: each degree of freedom is equally weighted, i.e. has the same probability of being emitted (thus colored particles are favoured with respect to the uncolored ones, high spin particles are favoured with respect to scalar ones, etc....). Furthermore, the emission is assumed to follow an adiabatic evolution: a homogeneous MBH temperature can be identified and slowly increases during the major part of the process of isotropic radiation.

\item {\it Planck phase}: in the final stages of evaporation when $ M_{MBH} \sim M_D$, the semi-classical and adiabatic approximation of General Relativity (which justifies a thermodynamical evolution of the evaporation process) breaks down and Quantum Gravity (QG) effects becomes much more important in defining the ultimate MBH fate. The possibilities range from a final remnant, an explosive break-up, or a complete evaporation, with each hypothesis still under discussion.
In particular, the role of discontinuous emissions with backreaction in this context must still be investigated. 
\end{itemize}

Each of the phases described above is subject to uncertainties. In particular,
a better understanding of the balding phase requires dynamical simulations that should also take into account the possible formation of exotic shapes (``saturn''-like configurations) or multiple MBH's immediately following the  collision. The democracy of the emissions that characterize the Schwarzschild phase is not present in the earlier phases where the MBH still retains a memory of the way in which it was created (it has hairs). In order to preserve unitarity it has recently been pointed out that democracy should be reached gradually, and that two scales should be introduced, instead of just one, to fully characterize the MBH evolution: in addition to the already mentioned gravitational scale/radius (at which gravitation becomes strong), a second (lower energy/higher distance) scale (e.g. the compactification radius in extra dimension models at which gravity deviations from the Einsteinian regime begin to manifest themselves), characterizes the transition from the non-democratic to the democratic emission regime \cite{dvali}. 
Furthermore, the emission of particles during MBH evolution is modified by gravitational effects, related to the curved geometry near the MBH horizon. 
These modifications are codified in grey-body factors and many results concerning their precise determination have recently appeared in the literature, thanks to increasingly more sophisticated computations. However, some of the factors are still unknown or very uncertain, like those for graviton emission in extra-dimensions from a rotating MBH. Finally, still unknown QG effects are expected to determine the evolution of the final MBH remnant in the Planck phase. While many works agree on the hypothesis of a complete evaporation, it is still possible that there may remain a finite MBH remnant.

Following progress in theoretical understanding, several numeric event generators have been developed in the last ten years for the simulation of MBH generation and decay, in particular {\texttt{Groke}}~\cite{ahn} in the framework of cosmic ray studies, and {\texttt{Charybdis}}~\cite{harris}, {\texttt{Catfish}}~\cite{cavaglia}, {\texttt{BlackMax}}~\cite{frost,dai} and {\texttt{QBH}}~\cite{gingrich} in the framework of LHC physics. 

The heavy particles (top quarks, Higgs and EW bosons, etc.) emitted by MBH evaporation decay quickly, i.e. before entering the detectors, and the partons and charged leptons emitted both by MBH evaporation and by these decays are further subject to parton and photon shower emissions, degrading their energy down to a scale where perturbative QCD can not be applied anymore, and hadronization takes place, followed by hadron decays. Non-perturbative effects in this context are described by means of phenomenological models.
This same chain of processes also occurs in p-p collisions in the framework of the SM and the corresponding physics and model parameters have been constrained over the years by results obtained at accelerators. In particular, shower Monte Carlo (SMC) programs like {\texttt{PYTHIA}}, {\texttt{HERWIG}} and {\texttt{SHERPA}} are commonly used to describe these processes\cite{buckley}. The largest uncertainties in this framework concern the complications that may arise when considering beams with a nuclear structure (as for instance in p-A and A-A collisions) and the propagation of the MBH decay products in a medium instead of the vacuum~\footnote{Actually, in case of p-p collisions medium effects reduces to the so-called ``underlying event'' effect, that is already one of the sources of the largest uncertainties in SMC simulations.}. 

Searches for MBH's have been conducted by the CMS and ATLAS experimental collaborations at LHC in the framework of the more general ``searches for exotica''. The analyses conducted so far~\cite{cms0,atlas0,cms1,cms2} have not lead to any evidence for MBH formation in p-p collisions at $E_{CM}$ = 7 TeV. However, these analyses have been criticized, since QG effects, expected to be important at LHC energies, have been neglected or treated too naively in the event generators used. Very recently, some theoretical work has also appeared in the literature pointing out, on the basis of other arguments like the generalized uncertainty principle or the extrapolation of the results of numerical simulation of colliding self-gravitating fluid objects, that the present LHC energy is in any case too low for the formation of MBH's~\cite{ali,rezzolla}. However, the situation is globally still controversial, and the exclusion at the present LHC energy certainly does not limit the possible formation of MBH's at higher energies. 

\begin{figure} 
\includegraphics[width=0.49\textwidth]{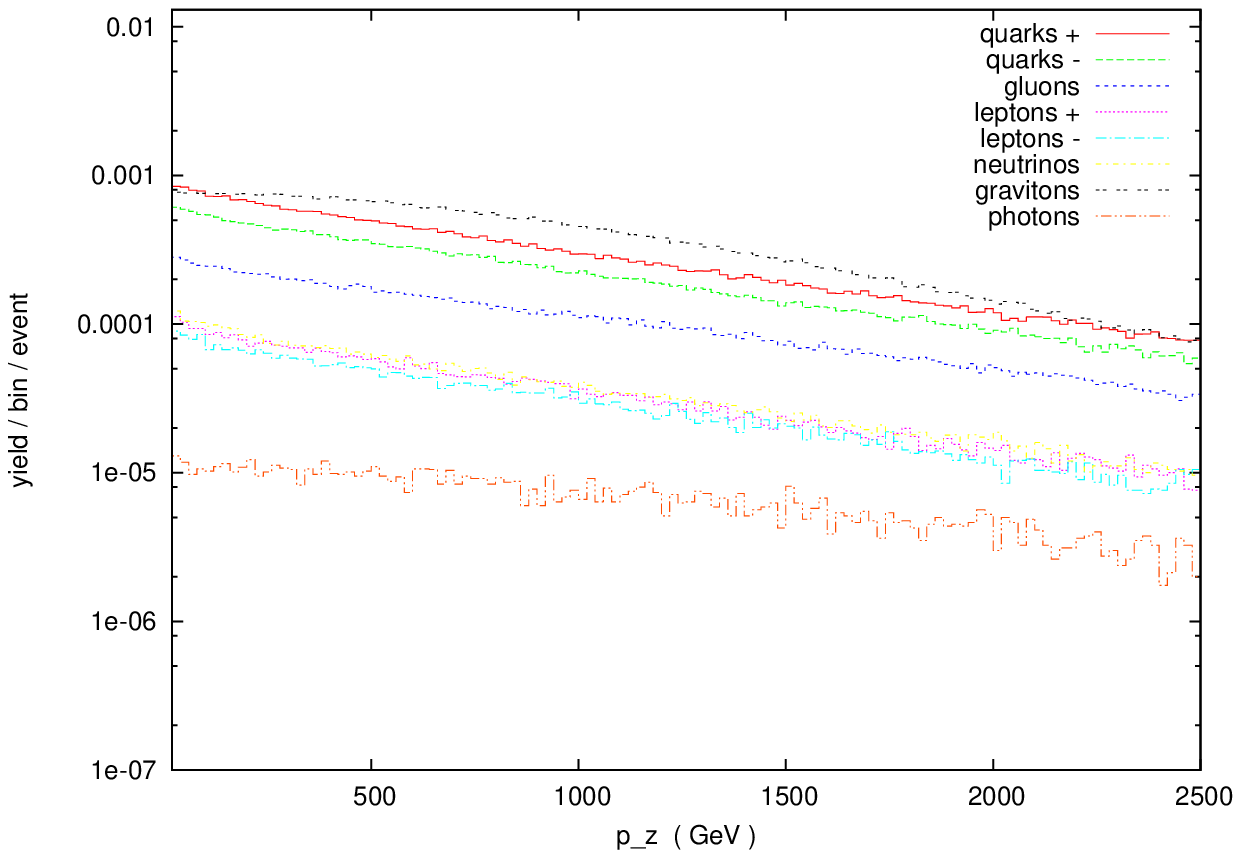}
\includegraphics[width=0.49\textwidth]{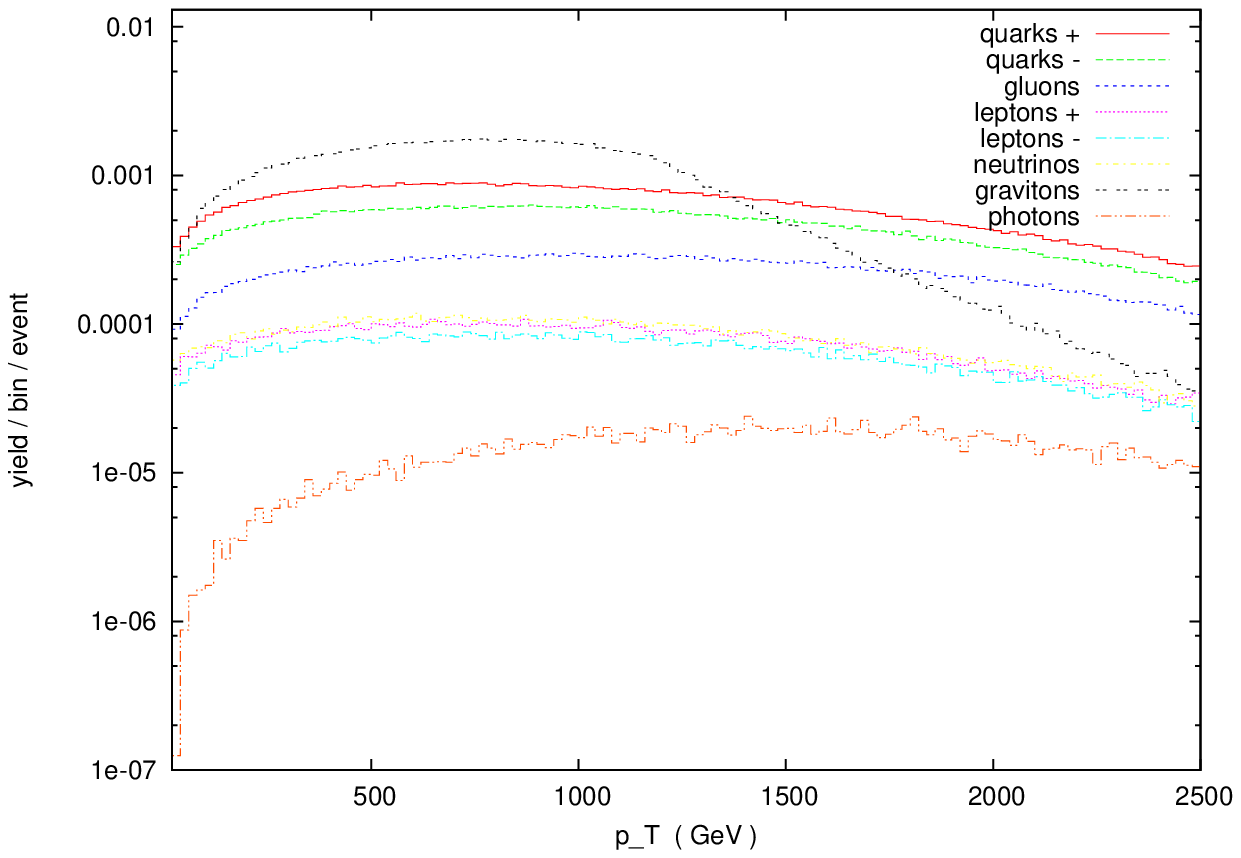}
\caption{\label{fig3}{Parallel ($left$) and transverse ($right$) momentum distributions for different SM degrees of freedom (quarks and antiquarks with positive charge, quarks and antiquarks with negative charge, gluons, positively charged leptons, negatively charged leptons, neutrinos, photons) and gravitons as computed by {\texttt{BlackMax}} for a MBH formed at a CM p-p collision energy $E_{CM}$ = 50 TeV for $M_D$ = 4 TeV. See text for more detail.}}
\end{figure}

\begin{figure} 
\includegraphics[width=0.49\textwidth]{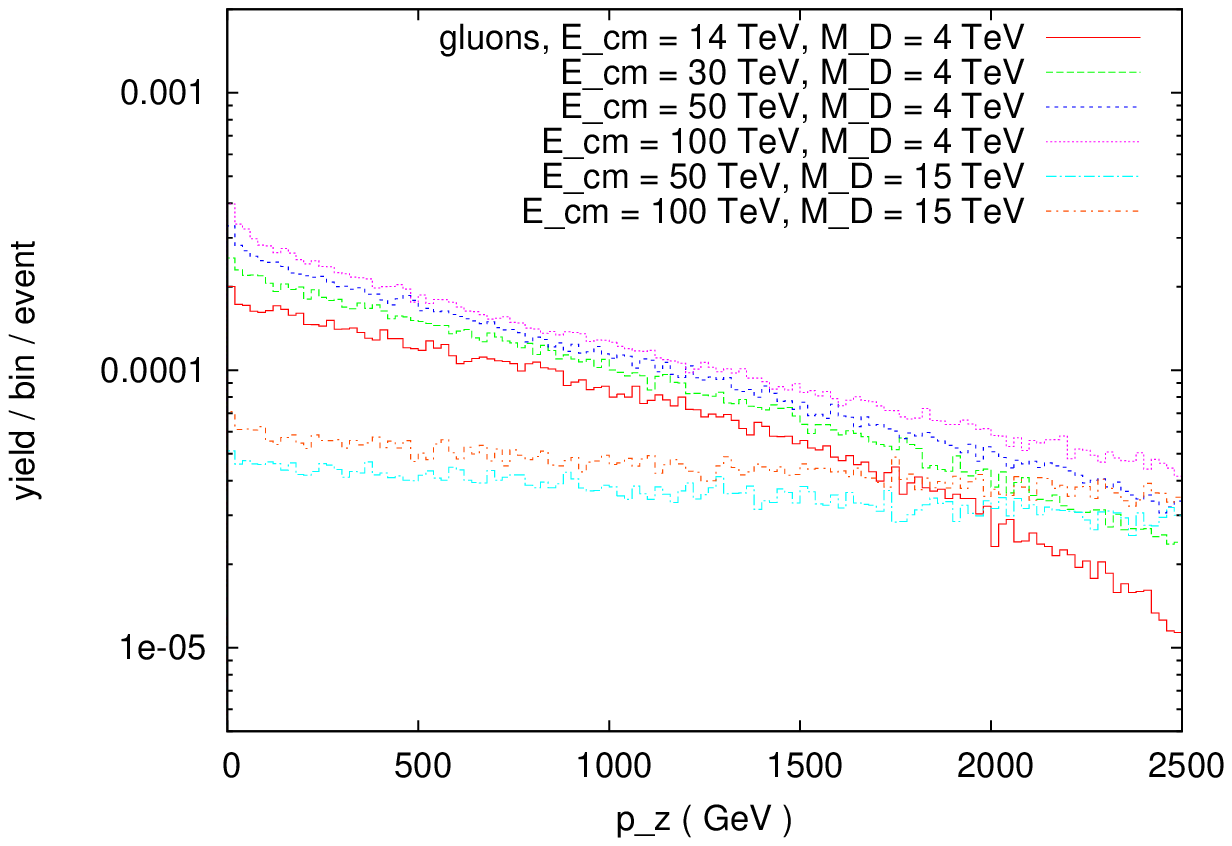}
\includegraphics[width=0.49\textwidth]{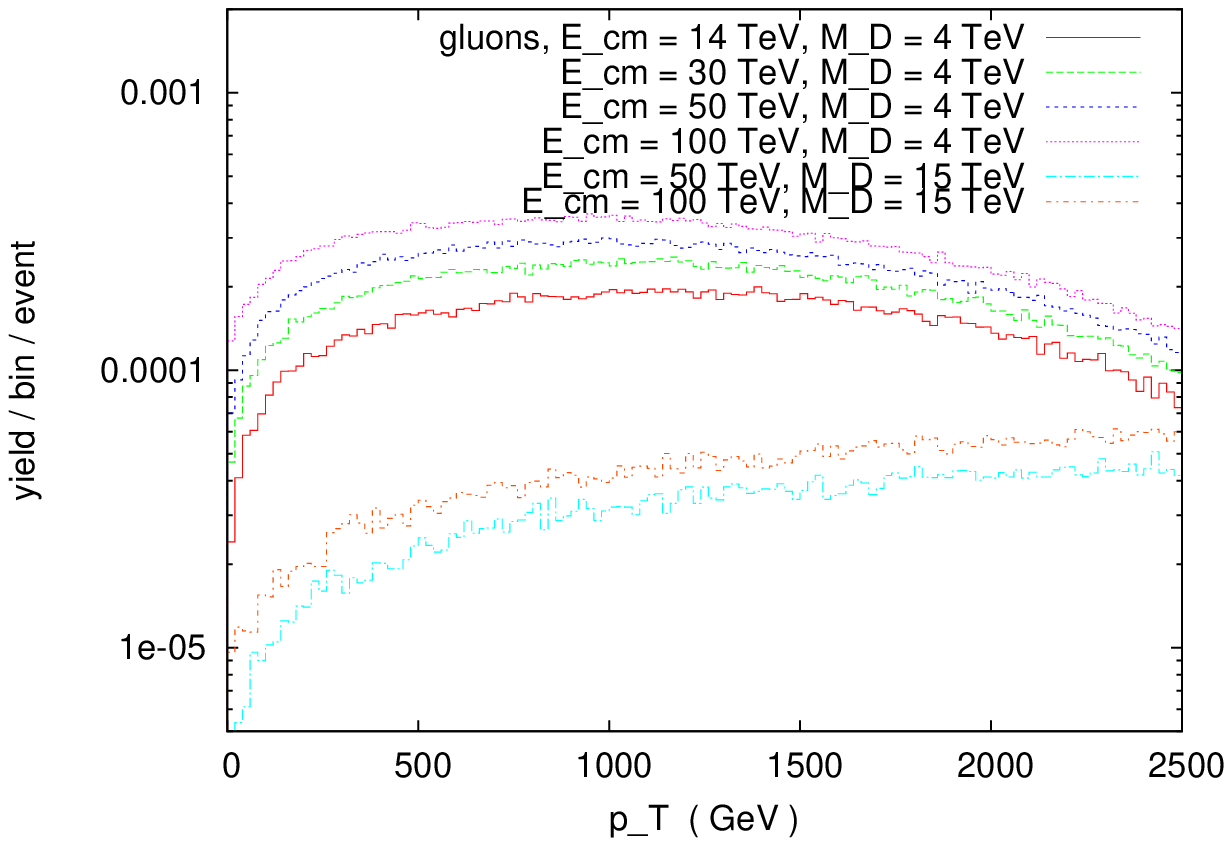}
\caption{\label{fig2}{Gluon parallel and transverse momentum distributions as computed by {\texttt{BlackMax}} for a MBH formed at four different CM p-p collision energies ($E_{CM}$ = 14, 28, 50, 100 TeV) for $M_D$ = 4 TeV and at two different CM energies ($E_{CM}$ = 50, 100 TeV) for $M_D$ = 15 TeV. See text for more detail.}}
\end{figure}

\begin{figure} 
\begin{center}
\includegraphics[width=0.49\textwidth]{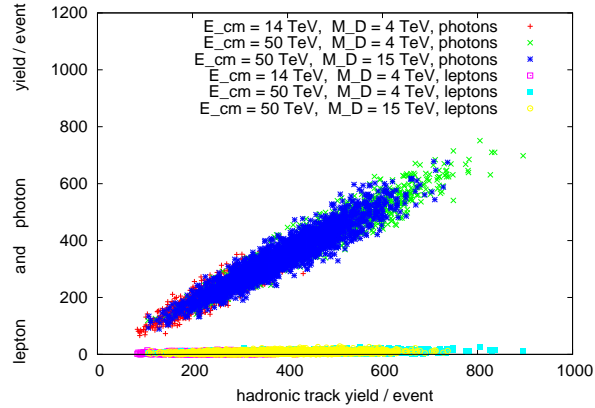}
\caption{\label{fig4}{Photon (upper part) and all lepton 
(lower part) yields as a function of the yield of all hadronic tracks after {\texttt{BlackMax + PYTHIA}}. Each point correspond to a different simulated event. Regions with different colors correspond to different $E_{CM}$ and $M_D$ parameters adopted in the MBH simulation, as labelled in the figure. See text for more detail.}}
\end{center}
\end{figure}

\begin{figure} 
\includegraphics[width=0.49\textwidth]{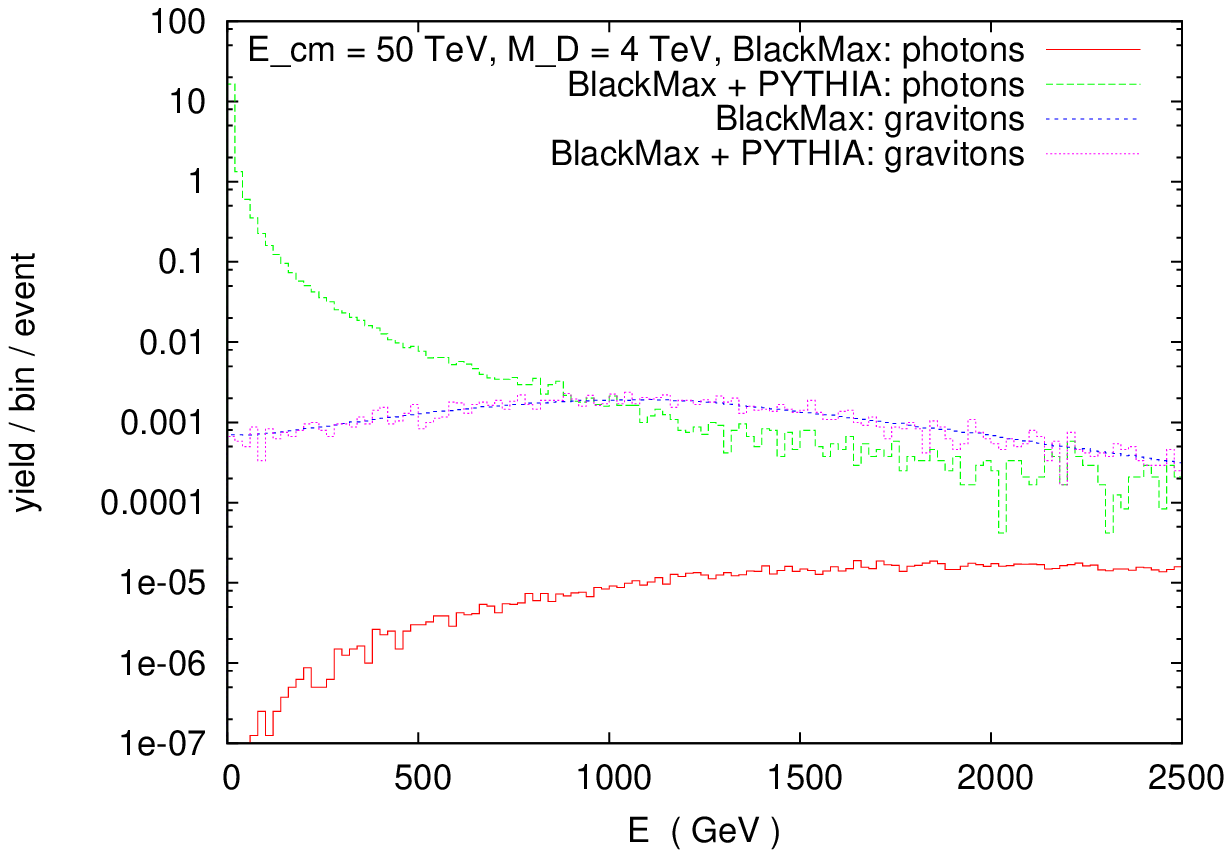}
\includegraphics[width=0.49\textwidth]{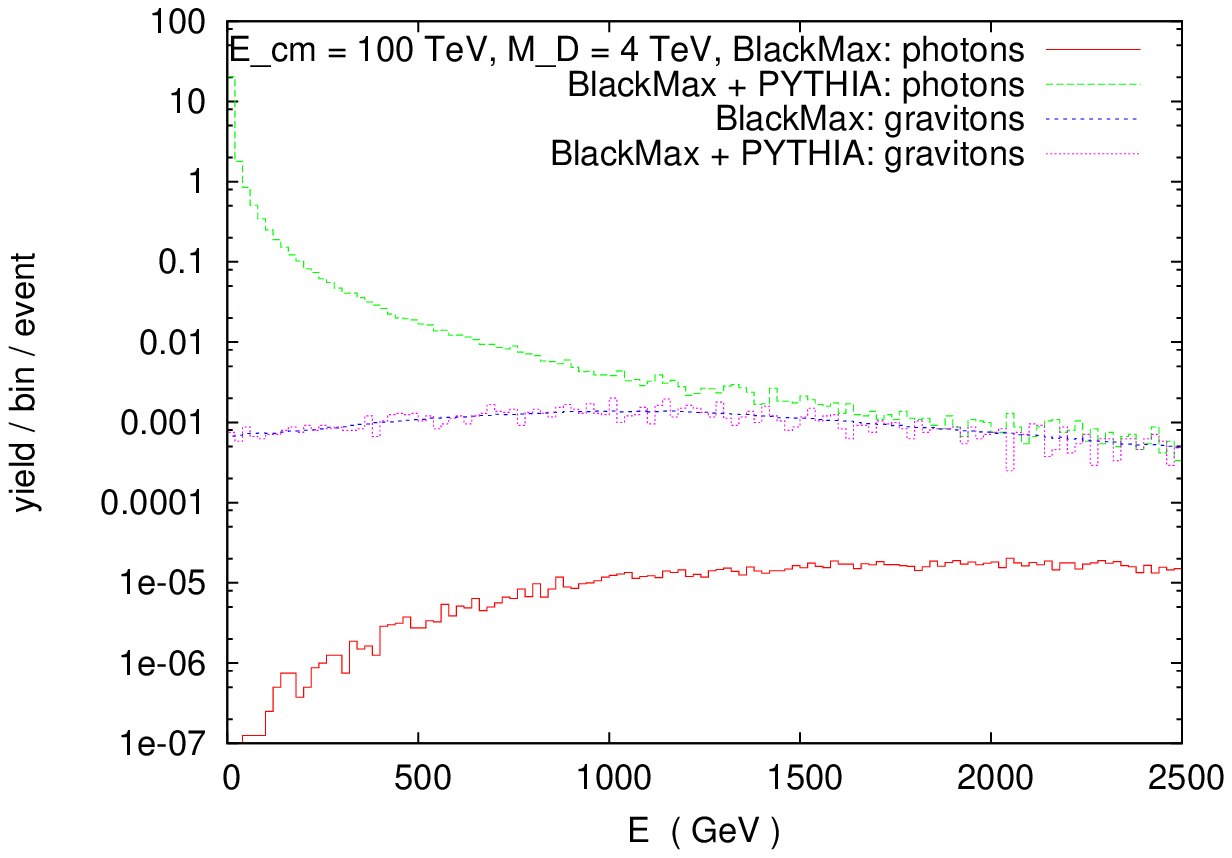}
\caption{\label{fig5}{Energy distributions of photons and gravitons
emitted by a MBH at a CM p-p collision energy $E_{CM}$ = 50 TeV ($left$)
and $E_{CM}$ = 100 TeV ($right$), for $M_D$ = 4 TeV. Results after both 
{\texttt{BlackMax}}
and {\texttt{BlackMax~+~PYTHIA}} are presented in each panel for comparison. See text for more detail.}}
\end{figure}

In this contribution we investigate the behaviour of event generators, usually adopted (and adapted) at LHC energies, at higher energies such as those reachable in the interactions of ultra-high-energy cosmic rays with the Earth's atmosphere, leading to extended air showers (EAS). In particular, we work with the last version of {\texttt{BlackMax}} (2.02.0), both in the standalone mode, and interfaced to the {\texttt{PYTHIA}} SMC code \cite{pythia}.  
We perform simulations of the formation of non-rotating MBH's in p-p collisions  in the 14 TeV $<$ $E_{CM}$ $<$ 100 TeV energy range, two different values for the fundamental gravity mass scale, i.e. $M_D = 4$ and 15 TeV, a MBH mass constrained in the range 2 $M_D < M_{MBH}  < E_{CM}$, and $n$ = 2 spatial extra dimensions without fermion splitting. In the simulation of MBH evolution, the mass, linear and angular momentum loss fractions were assumed to be equal to 0.3, whereas angular momentum, charge and color suppression factors were assumed to be equal to 0.2, and baryon and lepton numbers, as well as their difference, conserved. 
With these settings we investigated the kinematical properties of particles emitted during the MBH evolution as computed by {\texttt{BlackMax}} and also after the Parton Shower + Hadronization + Hadron decay chain, as computed by the interface of {\texttt{BlackMax}} with {\texttt{PYTHIA}}. Examples of selected results are presented in Figs.\ref{fig3}, \ref{fig2}, \ref{fig4} and \ref{fig5}.

In Fig.\ref{fig3}.a and \ref{fig3}.b the longitudinal and transverse momentum distributions (expressed in terms of number of particles/bin/event) are shown for different SM particle species for the case of MBH production at $E_{CM} =$ 50 TeV. After evaporation of the MBH the (anti-)quarks give rise to the largest contributions followed by gluons, (anti-)leptons and photons. Contributions from particles with opposite charges are shown separately: for any given flavour the contribution of positively charged particles is larger than that coming from negatively charged particles, probably due to the fact that during the final burst in the MBH evolution ({\texttt{BlackMax}} implements the hypothesis of complete evaporation), positive charged particles are predominantly emitted, because the majority of MBH's are positively charged (see also Ref.\cite{dai} for similar conclusions in a lower energy p-p study). 
The $p_z$ distributions are almost monotonically decreasing with similar slopes for all SM particles, whereas the $p_T$ distributions show some broad peaks, located at different $p_T$ values according to the particle species. (Anti-)leptons are emitted in pairs, i.e. as $\ell\nu_\ell$, $\ell^+ \ell^-$ or $\nu_\ell \bar{\nu}_\ell$, due to imposed lepton number conservation.
Graviton distributions are also shown, and display a high $p_T$ profile with a slope that decreases more rapidly than do those for SM particles, leading to a suppression of gravitons with respect to SM degrees of freedom at high $p_T$.

In Fig.\ref{fig2}.a and Fig.\ref{fig2}.b, the $p_z$ and $p_T$ distributions of a specific particle species, i.e. the gluon in this example, are shown as a function of the p-p collision $E_{CM}$ (leading to the formation of a MBH), for different values of $M_D$. It is evident that, for a fixed value of $M_D$, the shape of the distributions at different $E_{CM}$'s is preserved with the total number of gluons increasing with $E_{CM}$. This is as expected because the cross-section for MBH formation increases with $E_{CM}$. On the other hand, changing the value of $M_D$ leads to distributions with different shapes in addition to a changing value of the total cross-section. In particular, the position of the $p_T$ maximum for gluon emission increases with $M_D$, ranging from $p_T \sim$ 1.1 TeV for $M_D$ = 4 TeV to 
$p_T \sim$ 4.3 TeV for $M_D$ = 15 TeV.

The SM yields from MBH evaporation are in general modified after parton and photon shower + hadronization + hadron decay, as simulated by SMC codes, such as {\texttt{PYTHIA}}, which leads to hundreds of hadrons and photons. In particular, the number of emitted photons in each event turns out to be correlated to the number of emitted hadronic tracks, with a constant slope at increasing $E_{CM}$, as shown in Fig.\ref{fig4}. This slope is also independent of $M_D$, at a fixed $E_{CM}$. On the other hand, the total yield of emitted leptons turns out to be small (a few tens of particles) and does not show evident correlations with the number of hadronic tracks. This points towards the conclusion that the large number of photons is probably due light hadron (in particular $\pi^0$) decays, whereas electromagnetic shower effects are suppressed.

It is also interesting to compare the shapes of particle spectra at different stages of the evolution of the entire system. In particular this can be carried out for distributions of particles that are not subject to hadronization, such as leptons, photons and gravitons. In Fig.\ref{fig5}.a and \ref{fig5}.b the energy distributions of photons and gravitons at the parton level after MBH evaporation and at the hadron level after {\texttt{PYTHIA}} are shown, for two different $E_{CM}$ energies. It is evident that the contributions of the parton shower, the hadronization and hadron decay lead to a complete distorsion of the original photon spectrum, disproportionately populating the region of low energies with photons emitted in these last processes. The photon distributions at the evaporation level are very similar for both $E_{CM} = 50$ and 100 TeV, whereas, at the hadron level, the photon distribution at $E_{CM} = 100$ TeV is clearly much more populated than the corresponding one for $E_{CM} = 50$ TeV due to the stronger SMC effects. On the other hand, the graviton distributions are completely unaffected by shower effects, and in the case of $E_{CM} = 100$ TeV display a flatter profile in comparison to that at $E_{CM} = 50$ TeV. 

In conclusion, we have provided some examples of theoretical simulations demonstrating how parton shower + hadronization + hadron decay effects may dramatically modify particle distributions after MBH evaporation, especially in the case of some SM quanta, such as photons. This is certainly a challenge that must be confronted when trying to distinguish the effects of different MBH models, potentially observable through MBH formation, evaporation and decay in high-energy and ultra-high-energy collisions, such as those explored at LHC and in cosmic ray experiments. From our preliminary investigations it appears that lepton distributions are less affected than photon ones and should thus be preferred for these MBH studies.

\section*{Acknowledgments}
This work was financed by the Slovenian Research Agency (ARRS) and by the Slovenian Ministery of Work, under the AD-FUTURA program. We wish to thank the organizers of the 13$^{th}$ International Conference on Nuclear Reaction Mechanisms for financial support and for stimulating discussions.









\end{document}